# QUANTITATIVE INVESTMENT DIVERSIFICATION STRATEGIES VIA VARIOUS RISK MODELS

Maysam Khodayari Gharanchaei, Carnegie Mellon University, Pittsburgh, Pennsylvania, US
Prabhu Prasad Panda, Carnegie Mellon University, Pittsburgh, Pennsylvania, US
Xilin Chen, Carnegie Mellon University, Pittsburgh, Pennsylvania, US

**ABSTRACT**

*This paper focuses on the developing of high-dimensional risk models to construct portfolios of securities in the US stock exchange. Investors seek to gain the highest profits and lowest risk in capital markets. We have developed various risk models and for each model different investment strategies are tested. Out of sample tests are performed on a long-term horizon from 1970 until 2023.*

## 1. INTRODUCTION

The primary objective of this research is to assess and compare the investment features (Albuquerque et al., 2016) (Barberis et al., 2015) of various portfolio optimization approaches (Barillas and Shanken, 2018) (Barra, 1998) (Carhart, 1997) (Chamberlain and Rothschild, 1982) employing different covariance matrices (Albuquerque et al., 2016) (Chamberlain and Rothschild, 1982) (Cochrane, 2011). The focus is on delivering annualized measures of investments to meet investor preferences (Connor and Korajczyk, 1986). Key metrics (Connor and Korajczyk, 1988) (Daniel and Titman, 1997) considered include portfolio average excess return, annual standard deviation, annual Sharpe ratio (Connor and Korajczyk, 1986), compound rate, market beta coefficient, average positions over time, and the number of effective stocks (Daniel and Titman, 2011) (Ferson and Harvey, 1999) in the portfolio. All portfolios in this project are characterized as long-only and fully invested (Barberis et al., 2015) (Fama and French, 1993).

The study involves the construction of minimum variance, maximum diversification (Fama and French, 1993) (Fama and MacBeth, 1973), and risk parity portfolios (Fama and French, 2015) of n risky US assets using the CRSP dataset. Various covariance estimators, including single-factor covariance matrix, constant correlation covariance matrix, and sample covariance matrix with shrinkage (Fama and French, 2015) (Fan et al., 2016), were applied. Adjusted returns (Ferson and Harvey, 1999), excluding the effects of dividends, were used to enhance data accuracy (Freyberger et al., 2020). Hence, nine unique combinations of risk models and portfolio construction methods were compared against equally weighted and value-weighted portfolios, serving as market benchmarks.

Using single-factor and constant correlation covariance matrices (Gibbons et al., 1989) (Hansen and Richard, 1987) as risk estimators, the study revealed that the maximum diversification portfolio consistently outperformed other portfolio strategies. This superior performance is notably reflected in the Sharpe ratio. The portfolios constructed with the sample covariance estimator with shrinkage exhibited similar patterns, although performance differences were less pronounced compared to single-factor (Hou et al., 2015) and constant correlation covariance matrices. Notably, the study observed challenges in constructing the risk parity portfolio using the sample covariance estimator with shrinkage. This raises intriguing questions for future research endeavors.

In conclusion, this research sheds light on the significance of covariance matrix selection in portfolio optimization. The findings emphasize the robustness of the maximum diversification portfolio strategy and highlight potential challenges in implementing risk parity portfolios under certain covariance estimation methods.

## 2. METHODS DEVELOPED

### 2.1. Data

The dataset utilized in this research is sourced from the CRSP dataset, focusing on risky US stocks. To ensure accurate assessments, adjusted returns were employed to eliminate the impact of dividends from the data (Kelly et al., 2020). Several out of sample tests are performed on monthly returns (Kozak et al., 2020) over two periods, from January 1970 to December 2000, and from January 1970 to December 2023. For each month, the timeseries (Kozak et al., 2018) of the past 60 monthly excess returns and market excess returns (Lewellen, 2014) are used to estimate the covariance matrix and optimal holdings (Lewellen et al., 2010) for each scenario. The following diagram shows how we used timeseries of data to perform out-of-sample tests.

**Fig. 1.** Timeseries and Out-of-Sample Test Data Domain

| Timeseries Used for Estimating the Covariance Matrix | | | | Excluded Data | OOST Data |
|---|---|---|---|---|---|
| $r_{t_{-60}}$ | ………. | $r_{t_{-2}}$ | $r_{t_{-1}}$ | $r_{t_0}$ | $r_{t_1}$ |

### 2.2. Covariance Estimators

Three distinct covariance estimators were employed to capture the risk characteristics of the selected assets.

**2.2.1. Single-Factor Covariance Matrix:** In this approach, the covariance matrix (V) is derived using a single-factor model.

$$\boldsymbol{V} = \sigma_f^2 \cdot \boldsymbol{b}\boldsymbol{b}^T + \boldsymbol{D} \qquad (1)$$

In equation (1) $\sigma_f^2$ is the variance of market returns as the single factor of the risk model. **b** is the adjusted vector of factor loadings for n assets and **D** is the adjusted diagonal matrix of idiosyncratic volatilities of assets. To adjust these matrices, we use the following formulas for shrinking **b** and **D** respectively.

$$\beta_i := \hat{\beta}_i + \frac{1}{3} \cdot (1 - \hat{\beta}_i) = \frac{2}{3} \cdot \hat{\beta}_i + \frac{1}{3} \qquad (2)$$

$$\log \omega_i := \log \hat{\omega}_i + \frac{1}{3} \cdot \left( \frac{1}{N} \sum_{j=1}^{N} \log \hat{\omega}_j - \log \hat{\omega}_i \right) = \frac{2}{3} \cdot \log \hat{\omega}_i + \frac{1}{3N} \cdot \sum_{j=1}^{N} \log \hat{\omega}_j \qquad (3)$$

The above choice of shrinkage coefficients is based on empirical observations of the relationship between the estimated and realized values of betas and idiosyncratic volatilities. Finally, $\sigma_f^2$ is estimated via the usual sample variance (without any shrinkage).

**2.2.2. Constant Correlation Covariance Matrix:** In this approach, the covariance matrix is of the form

$$\boldsymbol{V} = \rho \cdot \boldsymbol{\sigma}\boldsymbol{\sigma}^T + (1 - \rho) Diag(\boldsymbol{\sigma})^2 \qquad (4)$$

Here ρ is the (constant) correlation between any two different stock returns and σ is the vector of stock volatilities. To estimate ρ and σ we estimated ρ as the average of all sample correlations, $\rho_{ij}$. Also, we estimated each $\sigma_i^2$ via sample variances and then shrink the log volatilities 1/3 towards their average log volatilities. All of the above $\hat{\beta}_i$, $\hat{\omega}_i^2$, $\hat{\rho}_{ij}$, and $\hat{\sigma}_i^2$ are computed via matrix operations (*loadings2(R,$r_M$)* method) as explained in the description of Project 2. We have also provided the regression based estimator as *loadings(R, $r_M$)* method in the code which is not used for estimating these parameters.

**2.2.3. Sample Covariance Matrix with Shrinkage:** For the last risk model, we estimated the covariance matrix by shrinking the sample covariance matrix towards a structured target matrix. The method employed in this model is the same method used by (Clarke et al., 2006).

### 2.3. Portfolio Construction

Finally, after building the risk models, we developed three distinct optimization problems to create minimum variance, maximum diversification, and risk parity portfolios. All the optimization problems can are solved by CVXPY library available for Python coding language. This library can leverage different solvers to solve optimization problems such as 'ECOS', 'SCS', 'OSQP', and 'OSQP'. We used 'GUROBI' as the default method in CVXPY library (Garces, 2021).

**2.3.1. Minimum Variance Portfolio:** This portfolio aims to minimize the portfolio's overall risk, measured as the variance of its expected returns. The fundamental idea is to find the combination of holdings that, when combined in a portfolio, results in the lowest possible level of portfolio volatility. The holdings are the optimal solution of the following optimization problem.

$$\min_{X} \quad X^T V X \quad (5)$$
$$s \cdot t \cdot \quad \mathbf{1}^T X = 1$$
$$X \leq u$$
$$x_i \geq 0 \quad i = 1, 2, \cdots, n$$

Here, **u** is an upper band set to avoid unintentional portfolio concentration on a few stocks.

**2.3.2. Maximum Diversification Portfolio:** In constructing portfolios, to avoid focusing on risk only, there are several approaches to diversify the assets in a way to gain better portfolio returns while controlling the risk distribution. The Maximum Diversification Portfolio is a portfolio optimization strategy designed to achieve the highest level of diversification among a set of assets to gain exposure to higher risks versus a greater risk degree of freedom. By assuming

$$\mathrm{K} X = Z \quad (6)$$

the maximum diversification optimization problem for a long-only and fully-invested portfolio can be expressed as

$$\min_{Z} \quad Z^T V Z \quad (7)$$
$$s \cdot t \cdot \quad \sigma^T Z = 1$$
$$\mathbf{1}^T Z = \mathrm{K}$$
$$Z \leq \mathrm{K} u$$
$$\mathrm{K} \geq 0$$
$$z_i \geq 0 \quad i = 1, 2, \cdots, n$$

Here, Κ is an arbitrary coefficient that satisfies the equation below.

**2.3.3. Risk Parity Portfolio:** The primary objective of a risk parity strategy is to ensure that each asset contributes equally to the total portfolio risk. This means that assets with higher volatility will receive a lower weight, and those with lower volatility will receive a higher weight. In other words, we are trying to distribute portfolio holdings in a way that neither asset has more risk contribution in the portfolio. Unlike the previous optimization problems, risk parity model is numerically sensitive because of the logarithmic term in the objective function. To overcome this problem, we set an upper band, a positive integer that is big enough to prevent the optimization problem to become non-positive definite (NPD). In this project we set this band equal to five and all the out-of-sample tests were completed successfully. The risk parity portfolio holdings for a long-only and fully-invested condition can be estimated by solving the following

$$\min_{Y} \quad \frac{1}{2} Y^T V Y - \sum_{i=0}^{n} \log(y_i) \quad (8)$$

$$s \cdot t \cdot \quad Y \geq 0 \quad and \; Y \leq d$$

Y is defined as

$$X = \frac{1}{1^T Y} Y \quad (9)$$

We observed that when the covariance matrix is based on sample covariance shrinkage model, PSD condition is not satisfied in the risk parity optimization problem and it needs further covariance matrix adjustment which is beyond the domain of this project.

## 3. RESULTS

We performed out of sample tests for 1000 assets over two periods, from January 1970 to December 2000, and from January 1970 to December 2023. The former period is the same period Clarke, de Silva, and Thorley used in the paper. Similarly, we got similar results for majority of the elements. The difference between our work and their paper is that we used numerical optimization solvers to compute the holdings over time but they leveraged analytical approach to gain a closed-form solution of holdings. To increase the efficiency and decrease computational cost of heavy computations, we leveraged matrix calculations to estimate elements of the models such as $\hat{\beta}_i$, $\hat{\omega}_i^2$, $\hat{\rho}_{ij}$, and $\hat{\sigma}_i^2$ and shrinkage operations as well. In the following sections, the results of out of sample tests from January 1970 to December 2023 via various risk models are presented. We used the average positions and effective N defined by (Strongin et al., 2000) in the following exhibit. The effective number of stocks in the portfolio, can be interpreted as the number of stocks that could be equal-weighted to get the same level of stock-specific risk as occurs in the original portfolio (Clarke et al., 2004).

**Exhibit. 1.** Performance of Market-Based Risk Model Portfolios from 1970 to 2023

|  | Market (Value-Weighted) | Equal Weighted | Minimum Variance | Maximum Diversification | Risk Parity |
|---|---|---|---|---|---|
| Average Excess Return | 10.1% | 9.7% | 3.7% | 27.94% | 8.7% |
| Standard Deviation | 15.7% | 17.4% | 10.4% | 19.5% | 15.4% |
| Sharp Ratio | 0.64 | 0.56 | 0.36 | 0.58 | 0.56 |
| Compound Return | 8.6% | 7.9% | 1.7% | 26.1% | 7.1% |
| Market Beta | 1.00 | 1.06 | 0.23 | 0.58 | 0.92 |
| Average Positions | 1000.0 | 1000.0 | 63.9 | 61.9 | 1000.0 |
| Effective N | 152.3 | 1000.0 | 44.6 | 43.5 | 894.1 |

**Exhibit. 2.** Performance of Constant Correlation Risk Model Portfolios from 1970 to 2023

|  | Market (Value-Weighted) | Equal Weighted | Minimum Variance | Maximum Diversification | Risk Parity |
|---|---|---|---|---|---|
| Average Excess Return | 10.1% | 9.7% | -1.3% | 41.8% | 9.0% |
| Standard Deviation | 15.7% | 17.4% | 10.2% | 40.4% | 16.9% |
| Sharp Ratio | 0.64 | 0.56 | -0.13 | 1.04 | 0.54 |
| Compound Return | 8.6% | 7.9% | -- | 34.3% | 7.3% |
| Market Beta | 1.00 | 1.06 | 0.41 | 1.85 | 1.03 |
| Average Positions | 1000.0 | 1000.0 | 34.9 | 33.7 | 1000.0 |
| Effective N | 1000.0 | 152.3 | 33.9 | 33.6 | 990.8 |

**Exhibit. 3.** Performance of Sample Covariance (Shrinkage) Risk Model Portfolios from 1970 to 2023

|  | Market (Value-Weighted) | Equal Weighted | Minimum Variance | Maximum Diversification | Risk Parity |
|---|---|---|---|---|---|
| Average Excess Return | 9.7% | 10.5% | 2.8% | 25.1% | -- |
| Standard Deviation | 15.5% | 17.2% | 6.7% | 18.8% | -- |
| Sharp Ratio | 0.63 | 0.61 | 0.41 | 1.34 | -- |
| Compound Return | 8.4% | 8.9% | 1.1% | 23.4% | -- |
| Market Beta | 1.00 | 1.05 | 0.36 | 0.98 | -- |
| Average Positions | 1000.0 | 1000.0 | 129.6 | 54.9 | -- |
| Effective N | 151.7 | 1000.0 | 77.4 | 40.7 | -- |

## 4. CONCLUSION

In the pursuit of optimizing investment portfolios, our research has delved into the nuanced intricacies of covariance matrix selection, shedding light on essential considerations for investors and portfolio managers. The comparison of various portfolio optimization approaches has revealed compelling insights into the performance differentials and challenges associated with specific covariance estimators. The last page includes the plots of additive returns of portfolios based on various risk models against time period from 1970 to 2000.

The stark performance differentials observed across the portfolio strategies underscore the critical role played by covariance matrix selection. Notably, the maximum diversification portfolio emerged as a standout performer, consistently outpacing other strategies in performance measures. This robust performance signals its efficacy in navigating the complex landscape of risk and return trade-offs. Our exploration into different covariance estimators has unearthed nuanced variations in portfolio outcomes. While portfolios constructed using single-factor and constant correlation covariance matrices exhibited pronounced performance disparities, the utilization of the sample covariance matrix with shrinkage mitigated some of these differences. This nuanced observation prompts a deeper examination of the interplay between covariance matrix selection and portfolio dynamics.

A noteworthy revelation is the challenge encountered in constructing risk parity portfolios under the sample covariance estimator with shrinkage. This obstacle unveils a potential avenue for future research, prompting a closer examination of the intricacies involved in achieving risk parity when employing specific covariance estimation methods. In conclusion, our research not only contributes insights to the ongoing discourse in portfolio management but also sets the stage for future investigations into the intricacies of covariance matrix selection and the challenges associated with specific portfolio optimization strategies.

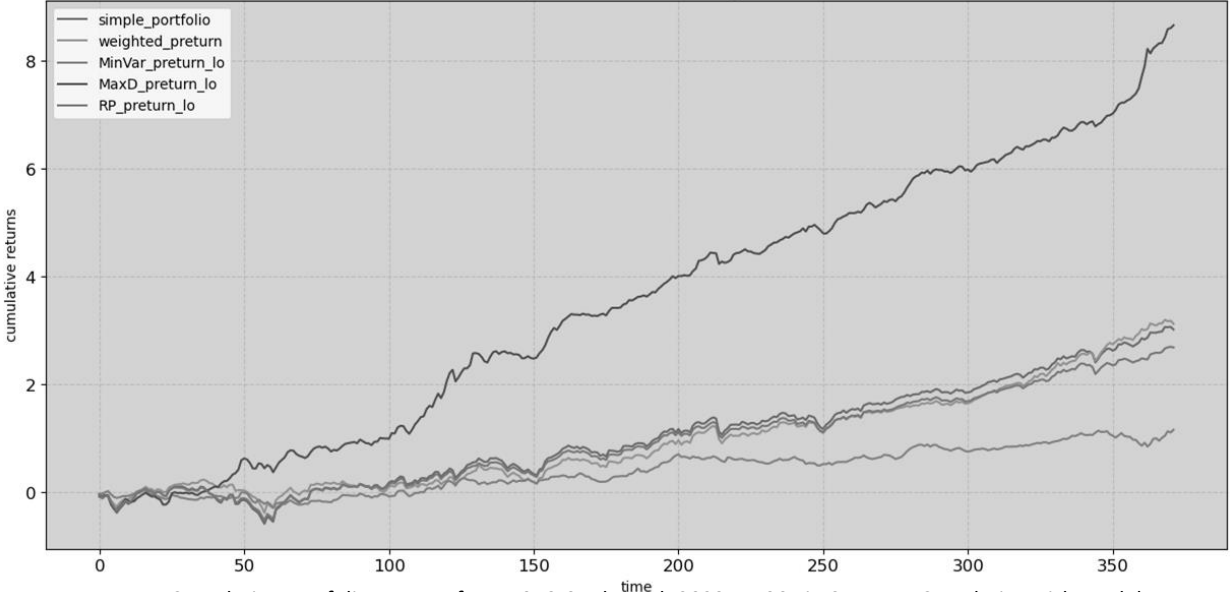

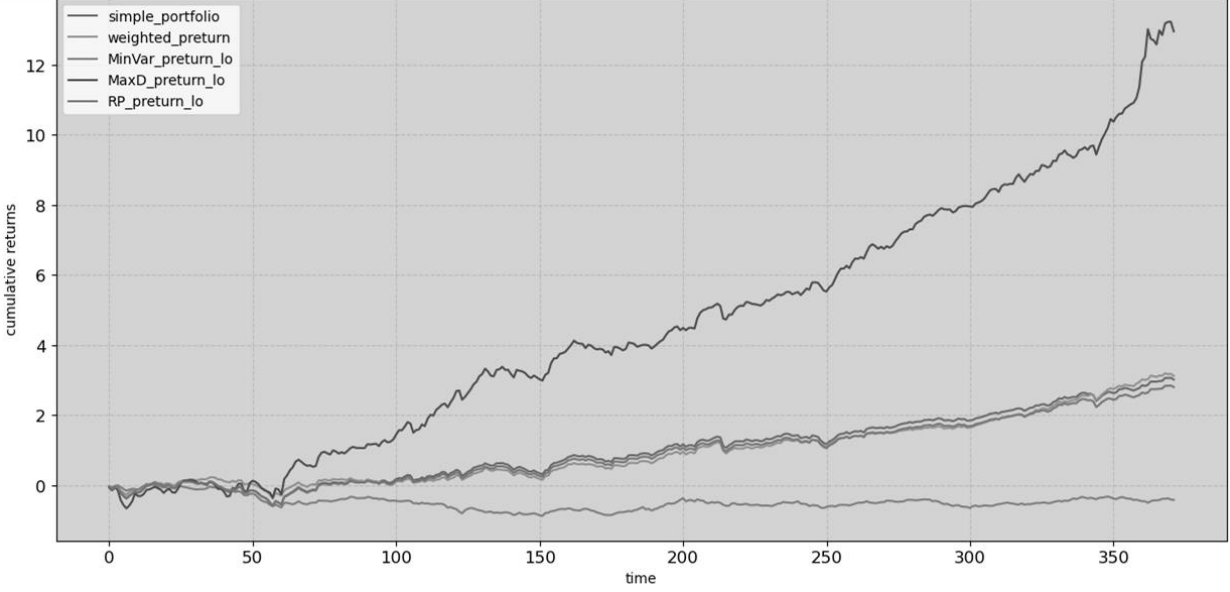

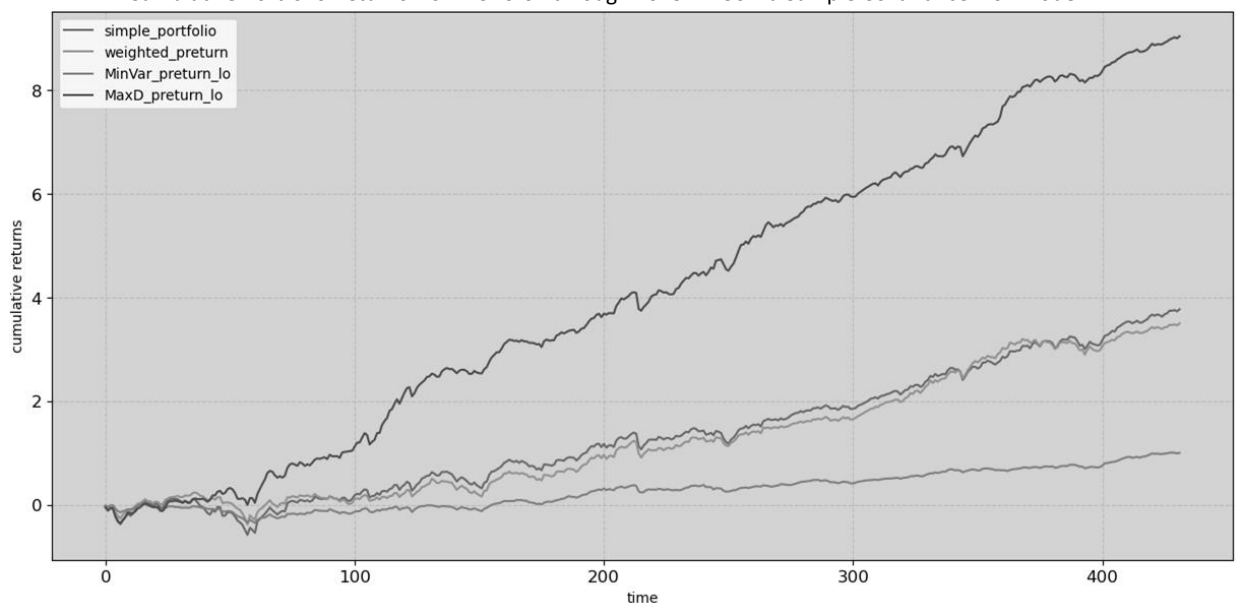

Cumulative Portfolio Returns from 1970-01 through 2023-11-30 via Sample Covariance Risk Model

## 5. References and Bibliography